\include{multicol}
\documentclass[letterpaper, 11.5pt]{paper}
\usepackage[superscript]{cite}
\usepackage{amssymb}
\usepackage{amsmath}
\usepackage{rotating}
\usepackage{comment}
\usepackage{enumerate}
\usepackage{subfigure}
\usepackage{lipsum}
\usepackage{color}
\usepackage{multirow}
\usepackage{geometry}
\usepackage{wrapfig}
\usepackage{comment}
\usepackage{mathtools}
\usepackage[super]{natbib}

\newgeometry{left=2.2cm,right=2.2cm, top=2.5cm, bottom=2.2cm}
\title{Penitentes as the origin of the bladed terrain of Tartarus Dorsa, Pluto}
\author{John E. Moores$^1$, Christina L. Smith$^1$, Anthony Toigo$^2$, and Scott Guzewich$^3$.}
\institution{$^1$ Centre for Research in Earth and Space Science, Department of Earth and Space Science and\\ \hspace{0.15cm} Engineering, York University, 4300 Keele Street, Toronto, Ontario M3J 1P3, Canada. \\ $^2$ Applied Physics Laboratory, Johns Hopkins University, Baltimore, Maryland, USA. \\ $^3$ NASA Goddard Space Flight Center, Greenbelt, Maryland, USA.\\ \\ \\ Published in Nature, Volume 541, p188-190, 12 January 2017.\\ DOI:10.1038/nature20779}

\begin{document}
\renewcommand*{\thefootnote}{\alph{footnote}}

\twocolumn[
\maketitle

\hrule

\begin{abstract}
Penitentes are ablative features observed in snow and ice that, on Earth, are characterized by regular cm to tens of cm spaced bowl-shaped depressions whose edges grade into tall spires up to several meters tall\cite{Nichols1939, Lliboutry1954, Claudin2015}. While penitentes have been suggested as an explanation for anomalous radar data on Europa\cite{Hobley2013}, hitherto no penitentes have been identified conclusively on other planetary bodies. Regular ridges with spacing of 3000 m to 5000 m and a depth of $\sim500$ m with morphologies that resemble penitentes (Fig. \ref{fig1}) have been observed by the New Horizons spacecraft\cite{Moore2016, Stern2015, Gladstone2016, Moore2017} in the Tartarus Dorsa (TD) region of Pluto (approximately 220-250$^\circ$ E, 0-20$^\circ$ N). Here we report simulations, based upon a recent model\cite{Claudin2015} adapted to conditions on Pluto\cite{Gladstone2016, Toigo2015}, that reproduce both the tri-modal orientation and the spacing of these features by deepening penitentes. These penitentes deepen by of order 1 cm per orbital cycle in the present era and grow only during periods of relatively high atmospheric pressure, suggesting a formation timescale of several tens of millions of years, consistent with cratering ages and the current atmospheric loss rate of methane. This time scale, in turn, implies that the penitentes formed from initial topographic variations of no more than a few 10s of meters, consistent with Pluto's youngest terrains.
\end{abstract}

\hrule

\bigskip

]

Conditions at Pluto are particularly suited to the creation of large penitentes (for more details on the theory of penitentes, please see the methods section and references\cite{Claudin2015, Cathles2014}). The Plutonian atmosphere is extremely stable, as evidenced by the presence of fogs in images and gravity waves consistent with wind speeds of 1 m s$^{-1}$ or less\cite{Gladstone2016} and as predicted by numerical models\cite{Toigo2015, Zalucha2012, Toigo2010} that argue for negligible horizontal surface wind speeds close to zero, never more than a few m s$^{-1}$ and often much less. Furthermore, the surface is composed of volatile ices, primarily nitrogen (N$_2$) and methane (CH$_4$) ice\cite{Stern2015}, close to their sublimation temperatures that are retained in the Plutonian system on long timescales due to cooling in the upper atmosphere\cite{Gladstone2016} that limits atmospheric escape. Finally, the low pressures observed by the New Horizons REX instrument\cite{Gladstone2016} of 1.1 Pa implies a long mean free path for molecular diffusion, increasing the thickness of the near-surface laminar sublayer, a thickness to which the spacing of penitentes is sensitive\cite{Claudin2015}.

\begin{figure*}
\centering
\includegraphics[trim=6.5cm 9cm 6cm 3cm, clip=true, width=0.5\textwidth]{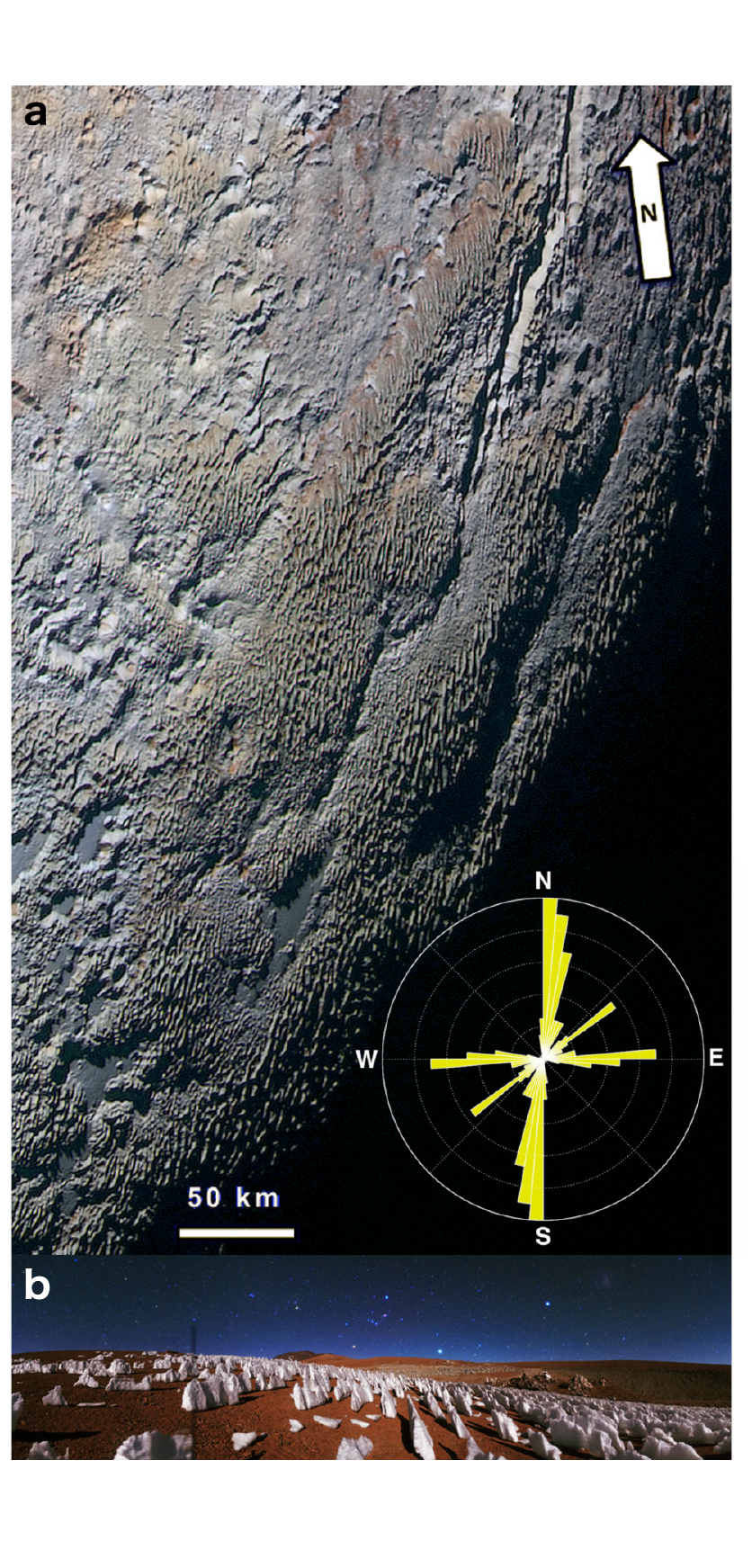}
\caption{Aligned ridges at Tartarus Dorsa, Pluto resemble high-latitude terrestrial penitentes. (a) Planetary Photojournal Image PIA19957 acquired with the Ralph/Multispectral Visual Imaging Camera (MVIC). Termed `bladed terrain,’ the region consists of ridged highlands. An examination of the geographic trend of 290 ridges covering 10,000 km$^2$ near the middle of the frame (rose diagram, inset) in a map-projected view\cite{Moore2016} indicates orientated features with three modes: north-south, east-west and northeast-southwest. Image Credit: NASA/Johns Hopkins University Applied Physics Laboratory/Southwest Research Institute. (b) Penitentes in the Atacama Desert showing aligned rows of blades oriented towards the mean sun direction. Image Credit: ESO/B. Tafreshi (twanight.org)}\label{fig1}
\end{figure*}

Using the ingress and egress values of the near-surface temperature as determined by REX\cite{Gladstone2016} of $37\pm3$ K and $45\pm3$ K and thermophysical data appropriate to N$_2$ and CH$_4$ at Pluto\cite{Brown1980, Young1997} we can determine whether penitentes are anticipated to form under present conditions and, if so, at what spacing. This data includes the saturated vapor pressures of CH$_4$ and N$_2$ as a function of temperature\cite{Brown1980, Young1997}, the thermal conductivity of N$_2$ and CH$_4$ ice, the latent heat of sublimation\cite{Brown1980} of CH$_4$ and N$_2$, the molecular diffusivity of CH$_4$ and N$_2$ in N$_2$ gas\cite{Hirschfelder1954} along with reasonable values of the surface albedo of 0.5, the sublimation kinetic rate constant of 0.4 m s$^{-1}$, and the depth of penetration of light into the ice of 1.6 cm, appropriate to a frosted scattering surface\cite{Claudin2015}. The value of the laminar sublayer thickness cannot be constrained directly and is therefore selected to reproduce the features seen at TD (see methods section for a discussion of model sensitivity). The reasonableness of this approach may be examined by considering friction velocity, a measure of the surface shear stress normalized by the fluid density, which is directly related to the laminar sub-layer thickness. Typically, the friction velocity is also a factor of a few smaller than the actual wind velocity.

\begin{figure*}
\centering
\includegraphics[trim=2cm 9cm 2cm 10cm, clip=true, width=0.75\textwidth]{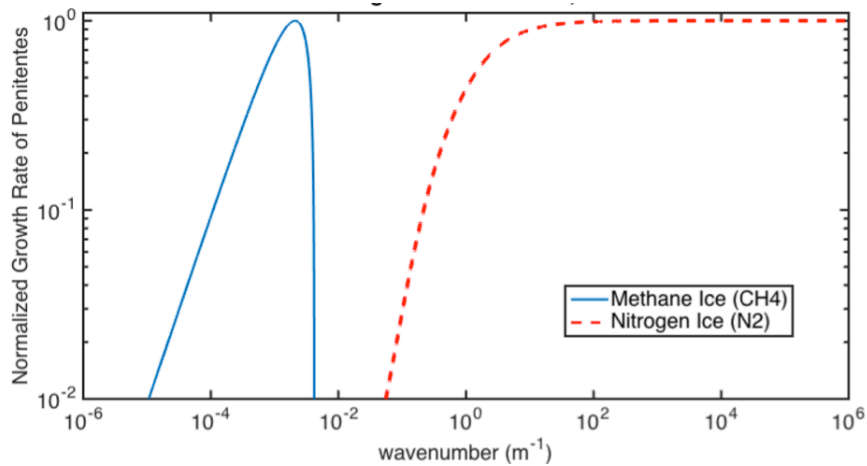}
\caption{Growth rates of penitentes of different spacings conditions appropriate for TD at the time of the New Horizons Encounter. A peak in growth rate is seen in methane ice, corresponding to a preferred penitente wavenumber of 0.0021 m$^{-1}$ (wavelength of 3000 m) when a laminar sublayer of 87 m is assumed. Nitrogen ice, shown here at the same laminar sublayer thickness, shows a monotonic increase in growth rate with increasing wavenumber, indicating that Nitrogen sublimates as a sheet. As predicted by this model, the penitentes observed by New Horizons occur only in terrains rich in methane ices.}\label{fig2}
\end{figure*}

Figure \ref{fig2} demonstrates that penitentes would be expected to form in regions rich in CH$_4$ ice, such as TD, but not in regions rich in N$_2$ ice under the conditions that prevailed during the New Horizons encounter. CH$_4$ ice is seen to have a peak in growth rate at a spacing consistent with the TD features provided that the laminar sublayer depth is approximately 87 m. This corresponds to wind velocities at the surface of a few cm s$^{-1}$. By contrast, N$_2$ ice is not seen to form penitentes for any value of the laminar sublayer thickness and instead the sublimation rate asymptotes to larger values at larger wavenumbers (smaller wavelengths), characteristic of sublimation as a sheet. 

Using the PlutoWRF modeling system\cite{Toigo2015}, with validation by the REX data\cite{Gladstone2016}, it is possible to extend the record of surface temperature and pressure over the entire Plutonian year. Pluto's orbital parameters, primarily the solar longitude of perihelion, the obliquity, semi-major axis and eccentricity may also be varied. Over long periods, greater than 20 Ma, these orbital parameters vary chaotically\cite{Sussman1988} between defined boundaries. Nevertheless, the fraction of time that Pluto spends in any one state can be estimated from previous orbital simulations\cite{Sussman1988,Applegate1986,Earle2015} and individual PlutoWRF simulations incorporating variations in orbital parameters were weighted accordingly. In addition to the atmospheric pressure, a primary factor affecting the orientation of growing penitentes, through the path of the sun in the sky at TD, is the sub-solar latitude at the time of perihelion, which is controlled primarily by the precession of perihelion. As such we chose to vary this parameter. The semi-major axis and eccentricity, which control Pluto's overall insolation, were left unchanged for all runs as they are currently close to their long-term mean\cite{Earle2015} and are not correlated with the value of the longitude of perihelion over long timescales. Obliquity, a relatively minor factor, was also left unchanged from its current value.

\begin{figure*}
\includegraphics[width=\textwidth]{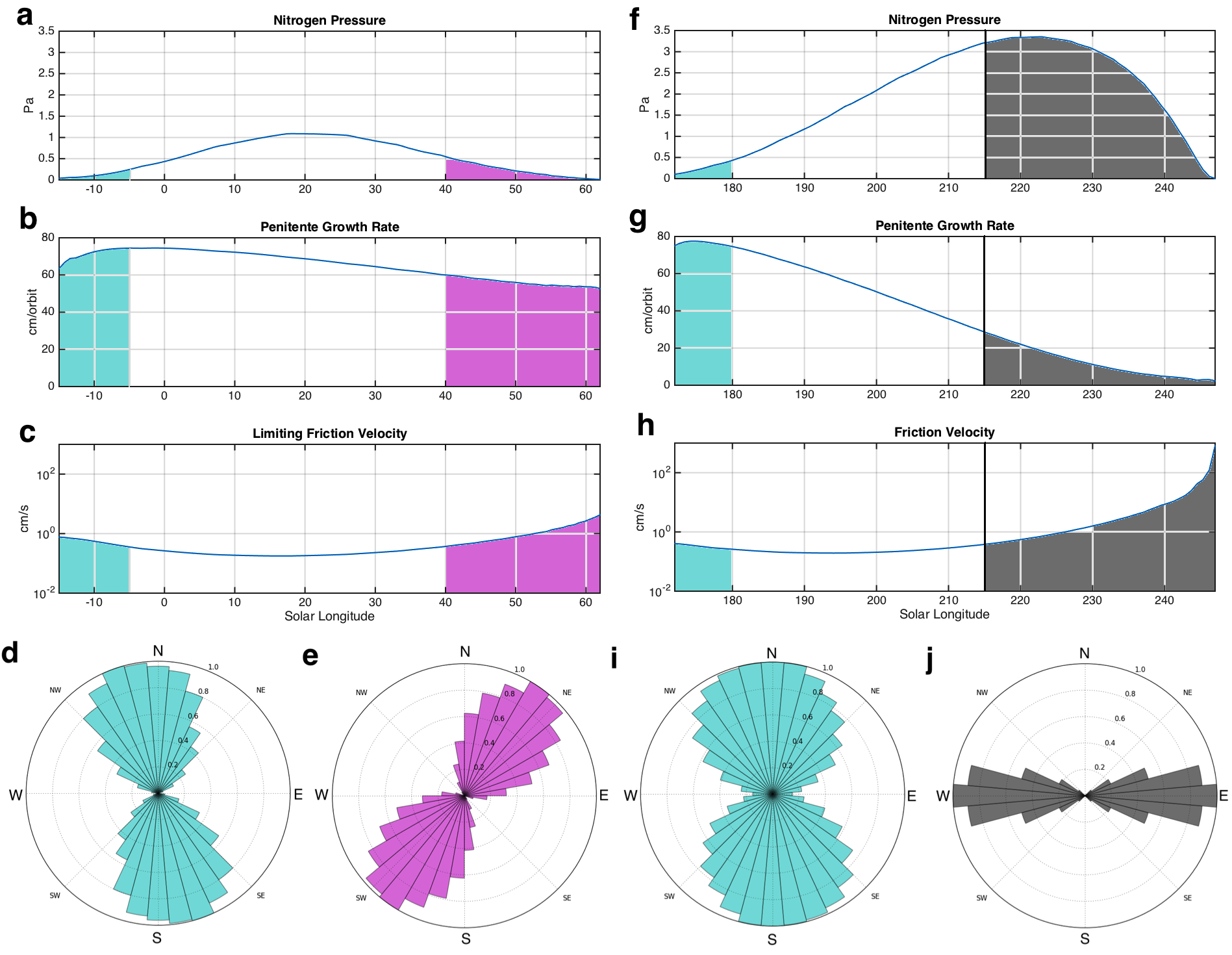}
\caption{Timing and orientation of penitentes formed, taking into account long-term variations in Pluto's orbit. Penitentes form only during one of two seasons when the atmospheric pressure is $>0.1$ Pa (top row). For the left column, a perihelion longitude of 325$^\circ$ is considered and, at right, 215$^\circ$. The second row indicates the model growth rate of the penitentes. As penitentes require calm winds to form it is reasonable to restrict the growth season to periods of higher limiting friction velocity (3rd row), indicated by shaded fill. The corresponding rose diagrams (bottom row) indicate the orientation of penitentes formed within each period.}\label{fig3}
\end{figure*}

Runs were produced with PlutoWRF for nine different values of the longitude of perihelion, simulating sub-solar latitudes at the time of perihelion varying from 60$^\circ$S to 60$^\circ$N. All simulations indicated an active period for the atmosphere either near the vernal equinox (L$_s$  = 0$^\circ$) or near the autumnal equinox (L$_s$  = 180$^\circ$) with the magnitude of the pressure peak corresponding to the alignment between the equinox and the longitude of perihelion. Outside of these periods, the atmosphere collapses, preventing a diffusive barrier to sublimating molecules from forming near the surface and short-circuiting the feedback required for penitente formation\cite{Claudin2015}. Figure \ref{fig3} shows the results of two representative simulations. The orbital parameters of both simulations resulted in a sub-solar latitude at the time of perihelion of 30$^\circ$S. In the plots on the left of Figure \ref{fig3} (panels a, b, and c), the sub-solar latitude is trending towards the north as L$_s$ progresses whereas the plots on the right (Fig 3f, g, and h) shows a southward trending sub-solar latitude with increasing L$_s$. As all nine of the simulations at different values of the longitude of perihelion produced a peak at one equinox or the other, each of the simulations shown in Figure \ref{fig3} is representative of approximately half of the long-term orbital states of Pluto, with the peak in pressure varying in between the limits of the two cases shown in Figure \ref{fig3}.

If penitentes are able to form at all times when the atmosphere is not collapsed, the average rate of penitente deepening over $> 20$ Ma timescales would be approximately 10 cm per orbit. However, such penitentes would not be strongly oriented. In particular, the strong east-west mode from Figure 1 can only form late in the active period of the autumnal equinox, during which time penitente growth rates are observed to be lower than at other times (Fig 3g and 3j). As such, penitente formation must be suppressed at many times of the active season. The suppression mechanism is suggested by previous work on the Earth, which has noted that low wind speeds are required for penitentes to form\cite{Claudin2015}. This suggests examining the limiting friction velocity; at larger values of the limiting friction velocity, larger wind-speeds are permitted before penitente growth is inhibited. 

If the formation of penitentes is restricted to times when the limiting friction velocity is relatively high, it becomes possible to identify four periods during which penitente formation is likely to be favoured. The highest friction velocities are found during the L$_s$ = 215$^\circ$ to 250$^\circ$ period when east-west oriented ridges are predicted to grow. Higher than typical friction velocity is also predicted early in both active periods when north-south oriented ridges are predicted, L$_s$ = 345$^\circ$ to 355$^\circ$ and 170$^\circ$ to 180$^\circ$. When the limiting friction velocity is relatively high, it is anticipated that penitentes may deepen during the entire day. However, as the friction velocity limit declines, penitente deepening likely becomes restricted to the late evening when winds would be anticipated to be the lightest. An asymmetry in scattered light around the limb of Pluto\cite{Gladstone2016} indicating an abundance of fog near the evening terminator strengthens the argument for low winds at this time. Further evidence for diurnal asymmetry in penitente formation is seen during the L$_s$ = 40$^\circ$ to 65$^\circ$ period, the only time of year when the northeast-southwest mode is possible and then only if the formation of penitentes is restricted to the late evening. The presence of this asymmetric mode at TD (Fig 1) argues strongly for a diurnal variation in the formation rate. 

Integrating the growth rate curves shown in Fig 3 we can place limits on the rate of deepening of these features in the current era. East-west penitentes may grow only from L$_s$ = 215$^\circ$ - 250$^\circ$, which means they can deepen at no more than $\sim1$ cm per orbit on average in the current era. North-south penitentes may grow at rates of up to $\sim2$ cm per orbit if they are able to grow during the entire day, however, the similarity between the friction velocity during the formation of the north-south and northeast-southwest penitentes suggests that north-south penitentes are likely to be diurnally restricted, growing no more than $\sim1$ cm per orbit in the current era. 

The rate of deepening of penitentes is not constant in time. As the features deepen, their growth rate accelerates as self-illumination becomes more effective\cite{Moore2016, Cathles2014}. Furthermore, as the ridges grow higher, the troughs are progressively shielded from the wind. This increases the fraction of the active season over which the penitentes can grow and helps to further promote the growth of favoured orientations. As such, deepening would have proceeded much less quickly in the past. This does not imply that there was less sublimation in the past; simply that such sublimation did not contribute to accentuating the relative topographic difference between the ridges and troughs of each blade. 

The current rate of growth of the features gives a formation time-scale of a few tens of millions of years to produce 500m of topography, assuming penitentes formed from initially quasi-random topographic variations with a scale of no more than a few tens of meters (see methods), perhaps comparable to the fields of pits\cite{Moore2016} observed on Pluto's youngest terrains\cite{Trowbridge2016, McKinnon2016}. This compares favourably to the cratering age of Tartarus Dorsa, whose single identified crater\cite{Moore2016} identifies it as intermediate in age between the relatively unmodified and adjacent Eastern Tombaugh Reggio (a billion years old)\cite{Moore2016} and Sputnik Planitia (less than 10 million years old)\cite{Moore2016}. The presence of penitentes on Pluto suggests that stable dynamic conditions and relatively high pressures are required\cite{Claudin2015} to create these features and suggests that they may form elsewhere in the solar system where the atmospheric conditions are appropriate.

\subsection*{Author Contributions}

This research was led by JEM who adapted the penitente model\cite{Claudin2015} to Pluto. CLS led the geometric modeling efforts. ADT and SDG provided valuable insights on the Plutonian atmosphere as well as output data from the PlutoWRF numerical atmospheric and surface energy balance models. JEM was supported in this work by a Discovery Grant (436252-2013) from the Natural Sciences and Engineering Research Council of Canada (NSERC) and CLS was supported by a fellowship under the Integrating Atmospheric Chemistry and Physics from the Earth to Space (IACPES) Collaborative Research and Training Experience (CREATE) program of NSERC.

\subsection*{Code Availability}

The code used to generate the plots shown in this paper and in the extended data set will be available at www.yorku.ca/jmoores/PlutoPenitentes.tar.gz upon publication.

\subsection*{Data Availability Statement}

The images of Tartarus Dorsa that support the findings of this study can be obtained from the NASA planetary photojournal (http://photojournal.jpl.nasa.gov) using the PIA identifiers noted in the Figure 1 caption. A map-projected view of the region can be found in reference 5, Figure 3B, and the orientations of the ridges derived from this map-projected view produced by JEM that were used to create the rose diagram of Figure 1a of this study will be available at www.yorku.ca/jmoores/ PlutoPenitentes.tar.gz upon publication. All other data used in the production of this paper are listed in the text and the figures were generated from this data using the code described under ‘Code Availability.’

\subsection*{Reprints and Permissions}

Reprints and permissions information is available at www.nature.com/reprints

\subsection*{Competing Financial Interests Statement}

The authors declare no competing financial interests.

\subsection*{Correspondence and Requests for materials}

Please address all correspondence and requests for materials, reprints, and permissions to {jmoores@yorku.ca}.

\bigskip
\bigskip

\hrule

\appendix

\section*{Extended Data}

\hrule

\bigskip
\bigskip

\renewcommand{\thefigure}{ED \arabic{figure}}
\renewcommand{\thetable}{ED \arabic{table}}

\setcounter{figure}{0}

\subsection*{Methods}

On the Earth, penitentes are most frequently observed in alpine equatorial environments where a combination of intense, nearly vertical illumination, low humidity and low wind speeds prevail\cite{Nichols1939, Lliboutry1954, Claudin2015}, however, penitentes at higher latitudes are observed to display trending linear crests aligned with the sun path rather than a purely cellular structure\cite{Nichols1939, Lliboutry1954, Cathles2014}. This arises due to the relatively low sun angle during penitente formation\cite{Cathles2014}, a geometry which is common on Pluto at many times of year for many latitudes due to the large obliquity (122.53$^\circ$) of the dwarf planet. As such, a successful identification of penitentes on Pluto must explain both the regular spacing and the orientation of the observed ridges while providing a formation timescale compatible with the inferred age of the terrain.

\subsubsection*{Dispersion Relation and the Theory of Penitentes}

It has been known for half a century\cite{Lliboutry1954} that penitentes grew from small initial depressions in a snow or ice surface that caused solar illumination to be focused on the center of the depression. This ‘self-illumination’ produced a progressively deepening topographic feature\cite{Cathles2014}, which eventually runs into other similarly growing features to produce the blade-like boundaries characteristic of penitentes. However, a field of penitentes, whether grown in the laboratory\cite{Bergeron2006} or observed in the natural environment\cite{Nichols1939, Lliboutry1954}, exhibits a regular cellular structure of a characteristic size that is typical of self-organizing systems. Self-illumination cannot explain this behavior as it is a scale-independent process that would be expected to lead to a wide variety of penitente sizes on a single surface, driven purely by the distribution of random initial topographic lows. 

Instead, the formation of penitentes and their regular spacing arises due to a balance between three factors\cite{Claudin2015}. Both (A) diffusion of heat in the ice or snowpack received from the sun and (B) self-illumination, the tendency of bowl-shaped depressions made of high albedo materials to concentrate light at the center of the depression, act to deepen penitentes by promoting sublimation. Opposing these actors is (C) molecular diffusion in the atmosphere above the penitentes in which an increase in relative humidity promotes condensation, slowing the rate at which penitentes deepen. As such, there tends to be a particular spacing of penitente which grows more quickly than other spacings, allowing a field of terrestrial penitentes to self-organize into the regular patterns that are observed (Fig 1). In addition to snowpack penitentes, this relationship has also been successful at describing regmaglypts\cite{Claudin2015}, the regular patterns observed in meteorite fusion crusts, that are formed much more quickly in different materials and at different scales than their icy counterparts. 

Under the influence of these three competing factors, a Mullins-Sekerka instability is created\cite{Mullins1964} which controls the characteristic size of the features produced. In the case of penitentes, a theoretical investigation of this instability mechanism\cite{Claudin2015} produced a dispersion relation for the growth rate, $\sigma$, reproduced with slight adaptation from recent work\cite{Claudin2015}:

\begin{equation}
\begin{multlined}
\sigma=\frac{\Psi_a}{\rho_S L} \left(\frac{k\Lambda}{1+P\tanh(kl)+Rk\Lambda} \right) \\ \shoveleft[1cm] \Bigg[ \left(1-\frac{1}{\sqrt(1-k^2\Lambda^2)}\right)+ \frac{\omega}{\pi}\left(1-\frac{k\Lambda}{\sqrt(1+k^2\Lambda^2)}\right)  \\ \shoveleft[4.5cm] -P\left(1-\frac{1}{\cosh(kl)}\right) \Bigg]
\end{multlined}
\end{equation}

\noindent Positive values of the growth rate are required to form penitentes. Here the three terms within the square brackets correspond to the three processes that affect the development of penitentes\cite{Claudin2015}. The first term corresponds to the effect of heat diffusion, the second term captures the self-illumination effect, and the third term captures the stabilizing effect of the molecular diffusivity of the atmosphere. The individual symbols in Equation 1 are listed in the table provided as extended data (Table ED 1) with their definitions and are chosen for consistency with previous work\cite{Claudin2015}. 

\begin{table*}
\centering
\caption{Summary of symbols used in the Penitentes Dispersion Relation (Equation 1).}
\includegraphics[trim=3cm 16cm 9cm 2cm, clip=true, width=0.75\textwidth]{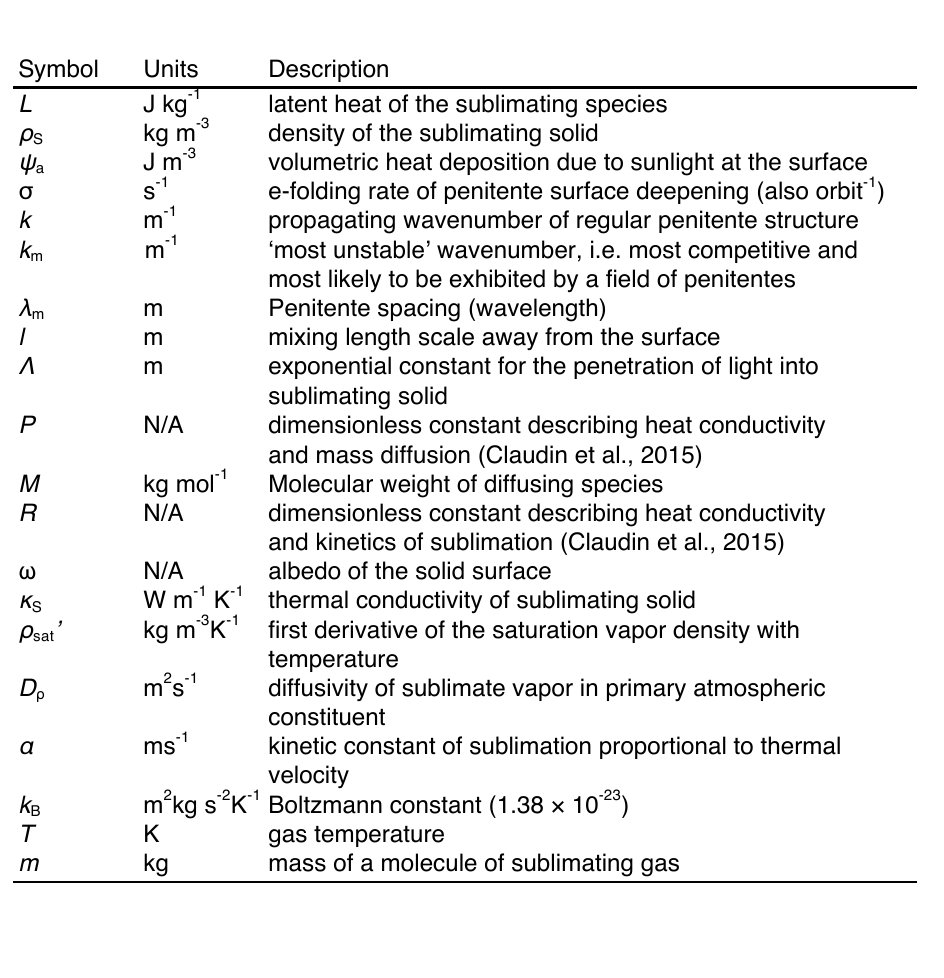}
\end{table*}

Two non-dimensional constants are employed to simplify Equation 1. These are P and R that compare the relative strengths of the heat conductivity and mass diffusion and the heat conductivity and sublimation kinetics, respectively, and are defined as\cite{Claudin2015}:

\begin{equation}
P=\frac{\kappa_S}{\rho'_{sat}D_\rho L}
\end{equation}

\begin{equation}
R=\frac{\kappa_S}{\rho'_{sat}\alpha\Lambda L}
\end{equation}

Finally, the spacing of the penitentes can be determined by calculation of $k_m$, the most unstable wavenumber, which is found by finding the wavenumber, $k$, that grows the fastest e.g. the maximum of $\sigma$, found by examining Equation 1 numerically. The regular spacing wavelength of the self-organizing cellular features is obtained from this most unstable wavenumber:

\begin{equation}
\lambda_m=\frac{2\pi}{k_m}
\end{equation}

\subsubsection*{Penitente Orientation}

To examine the effect of a non-vertical average sun vector, a model was prepared of parabolic cavities in otherwise flat terrain to examine the distribution of energy over the course of a single diurnal cycle at specific solar longitudes, comparable to previous terrestrial work\cite{Cathles2014}. To simulate the orientation offset from vertical of the penitentes, the parabola was manipulated with a rotational matrix: 

\begin{equation}
\begin{bmatrix}
    \cos(\omega)       & -\sin(\omega) \\
    \sin(\omega)       & \cos(\omega) \\
\end{bmatrix}
\end{equation}

\noindent where  is the anti-clockwise angle of rotation with respect to the vertical (see extended data, figure 1). This rotation was applied in both the positive and negative senses. At angles greater than 72$^\circ$, the opening of the penitente in the surface is too large to be practical, so these angles formed the limits in the positive and negative directions. The angle of rotation was increased in 5$^\circ$ increments between ±70$^\circ$. The grid resolution was kept constant, but due to the rotation, the number of points below the surface threshold varied with angle. 

A shadow may be caused from either or both edges of the penitente, depending upon the zenith angle of the incident flux and the rotation of the penitente. From either edge, a shadow is cast when the angle $\alpha$ is greater than $\theta$, shown in the hatched region in Figure ED 1. The regions in shadow are assumed to have zero flux incident as there is little atmospheric scattering on Pluto\cite{Gladstone2016}. The regions not in shadow receive a proportion of the incident radiation, corrected for the relative angles of incidence between the incoming radiation field and the normal surface vector:  

\begin{equation}
F_{rec}=\frac{-\vec{F}_{dir} \cdot \hat{n}}{|\vec{F}_{dir}|}
\end{equation}

\noindent where $F_{rec}$ is the received flux by the unit surface area, $F_{tot}$ is the magnitude of the flux field, $n$ is the normal surface vector and $F_{dir}$ is the vector dictating the direction of the flux field.

\begin{figure*}
\centering
\includegraphics[trim=2cm 20cm 2cm 20cm, clip=true, width=0.5\textwidth]{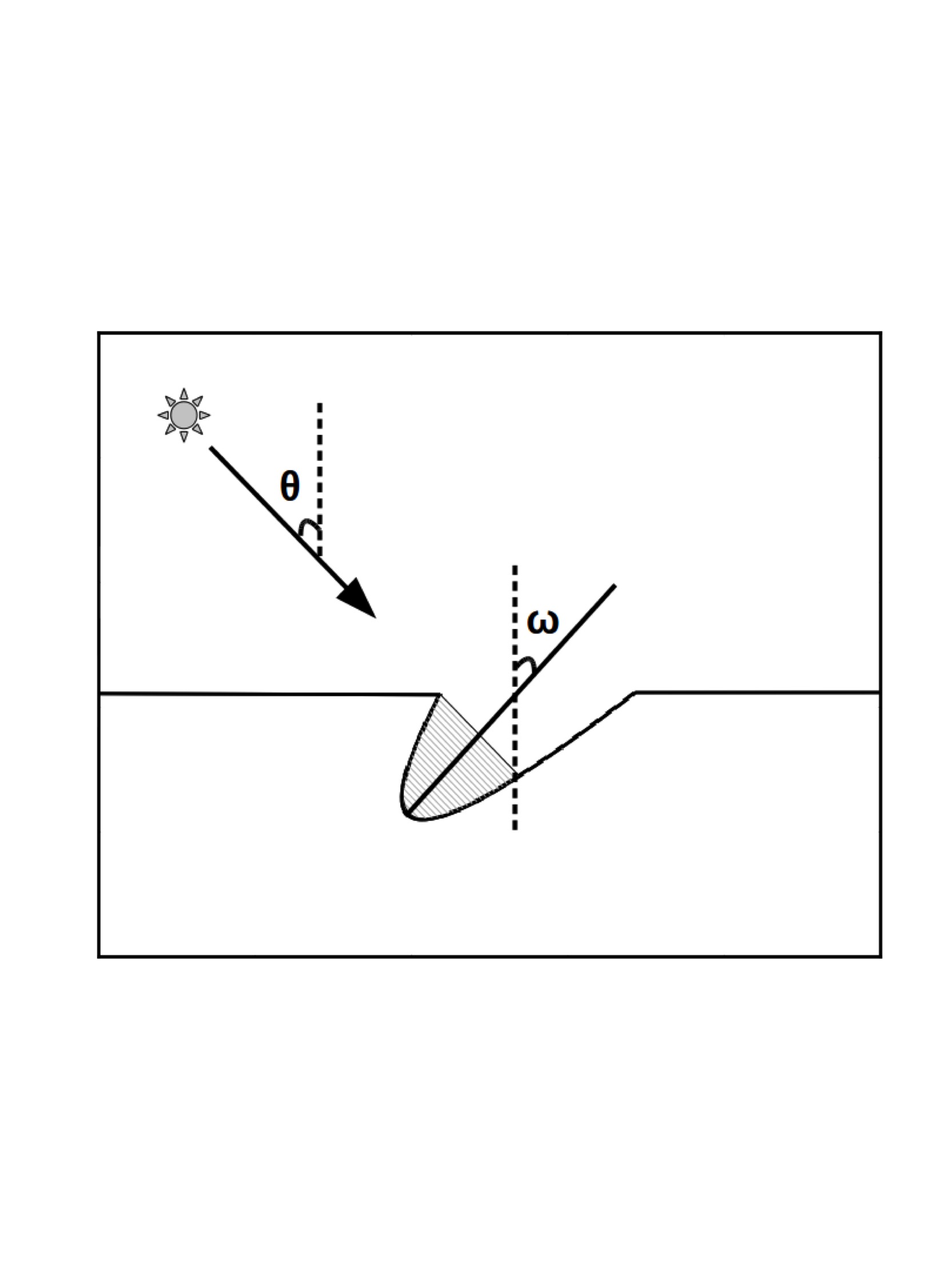}
\caption{Diagram showing the penitente geometry. Initially a parabolic depression in a flat, horizontal surface, model penitentes were generated by rotating by the angle $\omega$, with the positive sense indicated. The light grey region indicates the area in shadow under this orientation and projected solar zenith angle $\theta$, shown in the negative sense.}\label{ED1}
\end{figure*}

In order to calculate the total flux received on a specific date, the solar path across the sky at the site of the penitentes must first be calculated. This was done by adapting a well-validated martian model\cite{Allison2000} to Pluto, with 200 timing-points per diurnal cycle. The paths were calculated in 3-dimensions, specified by azimuth, $\phi$, and zenith, $\theta_Z$, angles. These angles were then projected into the plane of the penitente, resulting in a single projected zenith angle, $\theta$. The flux was reduced accordingly:

\begin{equation}
F_{proj}=F_{orig}\sqrt{\cos \theta_z^2+\left(\sin\theta_z\cos\phi\right)^2}
\end{equation}

\begin{equation}
\theta=\arctan\left( \frac{\cos\theta_z}{\sin\theta_z\cos\phi}\right)
\end{equation}

\noindent where $F_{proj}$ and $F_{orig}$ are the projected and original fluxes respectively.

The Solar paths and the incident Solar flux at the top of Pluto's atmosphere were computed at 5$^\circ$ L$_s$ intervals over the full 360$^\circ$ L$_s$ cycle. The instantaneous flux received by each grid point in the penitente was computed for each projected zenith angle and converted to an energy received. This was then summed over the entire penitente to give a total energy received and subsequently summed over the diurnal cycle.

Once the total flux received at each grid point was known, the contributions of self-illumination from other grid points within the parabolic structure were considered. The penitente surface was approximated as an isotropic (lambertian) scattering surface with the emitted energy proportional to the albedo multiplied by the received flux, as in comparable work\cite{Cathles2014}. The albedo was set at 0.5 for consistency with the dispersion relation model. Two scattering events were considered in each model run, which reduces the error as compared to a non-self illuminated model to less than 1 part in 8. An example penitente treated in this way for Ls = 230$^\circ$ can be seen in Figure ED 2. 

\begin{figure*}
\centering
\includegraphics[trim=2cm 18cm 2cm 18cm, clip=true, width=0.5\textwidth]{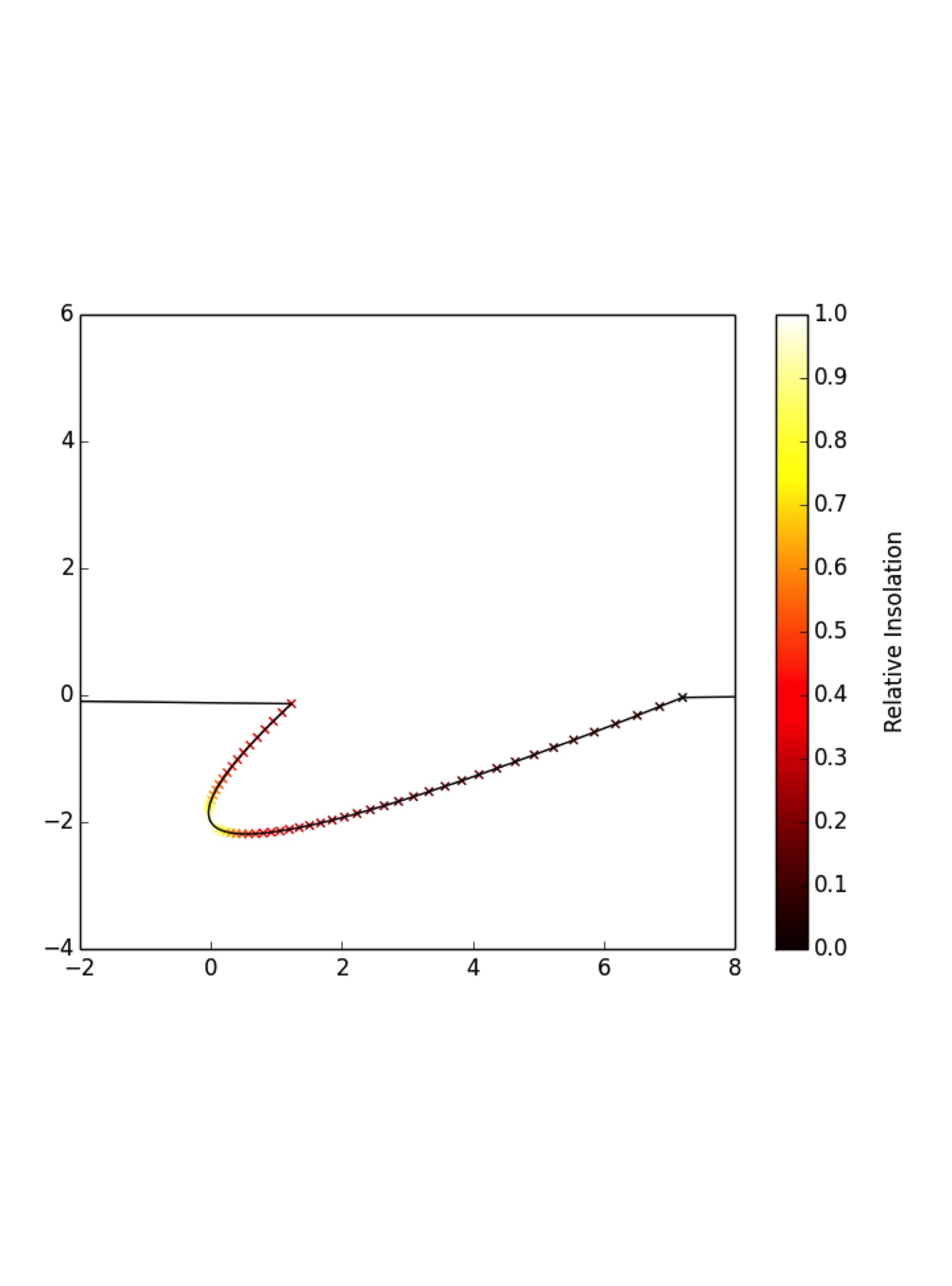}
\caption{A sample run of the penitente geometry for L$_s$ = 230$^\circ$ showing the distribution of received energy. This orientation of penitente at this time of year displays a concentration of energy near the bottom arising from orientation and self-illumination.}
\end{figure*}

Finally, the tilted penitentes were rotated in a horizontal plane to create the rose diagrams shown in Figure 3 with the energy received propagated exponentially in time, as per Equation 12. 

\subsubsection*{Penitentes in Methane Ice}

The values needed to calculate $P$ and $R$ for methane on Pluto for the season of the New Horizons encounter may be acquired from a combination of observation and laboratory experimentation. Data on the saturated vapor pressure are available\cite{Brown1980, Young1997} which suggests a value of between $2\times10^{-5}$ Pa to $8\times10^{-4}$ Pa depending on the fractional surface coverage of methane. The latent heat of sublimation of CH$_4$ at the relevant temperature is approximately\cite{Stansberry1996,Brown1980} $5.5\times10^5$ J kg$^{-1}$. The molecular diffusivity of methane in a binary mixture of nitrogen and methane can also be calculated using data on the van der Waals radius\cite{Hirschfelder1954} of the two molecules. The thermal conductivity of CH$_4$ ice can be calculated\cite{Lupo1980} and at these temperatures is 0.42 W m$^{-1}$ K$^{-1}$.

The value of $\rho'_{sat}$ is more challenging to obtain. While this can be calculated from the saturation pressure curves, a simpler method derived from the Clausius-Clapeyron equation has been used in other work\cite{Claudin2015} and for consistency is used here as well. 

\begin{equation}
\rho'_{sat}(T)=\frac{\rho_{sat}(T)}{T}\left(\frac{ML}{RT}-1\right)
\end{equation}

\noindent Given the values above, this suggests $\rho'_{sat}\sim2.5\times10^{-6}$ kg m$^{-3}$ K$^{-1}$.

Other parameters are more loosely constrained by the available data and as such, the sensitivity of even large changes in these variables to the solutions provided was evaluated as part of our analysis. In this category are terms such as $\alpha$, $\omega$, and $\Lambda$. 

Water ice under terrestrial conditions, near 273 K, was estimated to have a value for the sublimation kinetic rate constant, $\alpha$, of between 1 and 100 m s$^{-1}$ with 1 m s$^{-1}$ being used in calculations\cite{Claudin2015}. This constant is proportional to the thermal velocity of the sublimating molecules or:

\begin{equation}
\frac{\alpha_{CH_4}}{\alpha_{Ice}}=\sqrt{\frac{T_{CH_4}}{m_{CH_4}}\frac{m_{Ice}}{T_{Ice}}}
\end{equation}

\noindent Therefore for methane at Pluto surface temperatures, we suggest a typical value of 0.4 m s$^{-1}$ for $\alpha$.

The albedo of methane, $\omega$, is somewhat more difficult to constrain. Complicating matters is the tendency of methane ice to be converted photo-chemically into forms (tholins) with different colorations and albedos. For the present time, therefore, the albedo of methane ice will be taken as 0.5. It should be noted that albedos of 0.2-0.8, which cover the reasonable range of values for objects with methane ice on their surfaces, do not appreciably change the results. 

Finally, the penetration depth of light into ice, $\Lambda$, is a difficult parameter to estimate. While there is evidence to suggest that nitrogen ice may be particularly clear\cite{Brown1990}, the upper surface is likely composed of a large number of optical scattering centers (e.g. frost), which indicates that the value observed in terrestrial snow\cite{Claudin2015} of 0.016 m is likely also applicable to frosted methane with photochemical products preventing deep penetration. Nevertheless, larger values were also attempted. These did not change the value of the penitente spacing, but did strongly affect the growth rate of the penitentes with larger values of the penetration depth corresponding to more rapid incisement.

By contrast, solutions for the most unstable wavenumber were sensitive to the value of $l$, the laminar sub-layer thickness\cite{Claudin2015}. This result may be examined for stability to changes in $\alpha$, $\omega$ and $\Lambda$. For $\alpha$, a decrease by an order of magnitude, unlikely as we derive our figure from the bottom of the reasonable range\cite{Claudin2015}, increases the penitente wavelength by 7\%. Increasing $\alpha$ by two orders of magnitude decreases the penitente wavelength by 1\%. Varying $\omega$ by decreasing its value to 0.2 yeilds a 40\% increase in penitente wavelength. Increasing the value to 0.8 decreases the penitente wavelength by 14\%. Finally, increasing $\Lambda$ by two orders of magnitude to 1.6 m from 0.016 m increases the penitente wavelength by a 0.26\% whereas decreasing $\Lambda$ by several orders of magnitude has no appreciable effect.

\subsubsection*{Friction Velocity and Wind Speed}

The friction velocity can be related directly to the laminar sublayer thickness through\cite{Claudin2015,Melosh2011}:

\begin{equation}
u_*\approx\frac{5\mu}{\rho l}
\end{equation}

\noindent This relationship provides an upper limit on the friction velocity, given the size of the features observed. Given that the fluid viscosity, $\mu$, is approximately $2.4\times10^{-6}$ kg m$^{-1}$ s$^{-1}$ and the density\cite{Gladstone2016} at $11 \mu \rm{bar}$ (1.1 Pa) is approximately $9.3\times10^{-5}$ kg m$^{-3}$ we get a derived value of the friction velocity of a few cm s$^{-1}$. This implies a very low free-stream velocity above the features of at most a few tens of cm s$^{-1}$ based on planetary extrapolations of the Karmàn boundary layer model\cite{Melosh2011}. This result is surprisingly similar to what is observed for terrestrial penitentes. Here, similarly low wind conditions are required for formation\cite{Claudin2015} with morphologies transitioning to suncups and cross-hatched terrains at higher wind speeds. 

\subsubsection*{Penitentes in Nitrogen Ice?}

Corresponding values for nitrogen ice as compared to methane ice can also be calculated. We derive a value for $\rho'_{sat}\sim2\times10^{-3}$ kg m$^{-3}$ K$^{-1}$ using Equation 8.  The thermal conductivity of nitrogen ice is given as 0.2 W m$^{-1}$ K$^{-1}$ and the latent heat is approximately\cite{Stansberry1996,Brown1980} $2.5\times10^5$ J kg$^{-1}$. Analogous to our calculations for methane, we find that the self-diffusion of sublimating N$_2$ into an N$_2$ atmosphere is 0.108 m$^2$ s$^{-1}$. While larger values are perhaps appropriate for the albedo of N$_2$, as Triton has an albedo of 0.76, we use 0.5 for comparison with the results for methane. Furthermore, $\alpha$ should be approximately 0.3 m s$^{-1}$ for N$_2$ using Equation 9. We allow a larger value for $\Lambda$ of 1 m to account for the assumed clarity of nitrogen ice. Lastly, for consistency with methane, we assume the same laminar sub-layer thickness, which ultimately implies the same wind speed, though this was varied to see if any laminar layer value would produce penitentes in N$_2$ ice. No combination of reasonable parameters resulted in penitente formation in N$_2$ ice.

\subsubsection*{Seasonal Dependence}

The PlutoWRF model has been extensively discussed previously\cite{Toigo2015}. There were three components that were considered for this application. First, during the L$_s$ = 0$^\circ$-65$^\circ$ period, the atmosphere is sufficiently thick that the complete global circulation model (GCM) may be utilized. This model was validated against New Horizons REX data\cite{Gladstone2016}. For both ingress and egress, the GCM-reported values at the location and local time of the center of the observations matched the reported values to within the 3K error\cite{Gladstone2016}. For ingress at 17.0$^\circ$S, 16:31, the value reported by PlutoWRF was 39.7 K and for egress at 15.9$^\circ$N, 04:42, the value reported by PlutoWRF was 45.7K. On the average, the PlutoWRF values are warmer by 1.7 K and this slight adjustment was made to model temperatures. Furthermore, the pressure reported by PlutoWRF of 0.87 Pa is slightly low compared to the REX observation of 1.0-1.1 Pa, so a slight adjustment was also made to the pressures reported by PlutoWRF.

Outside of the L$_s$ = 0$^\circ$ - 65$^\circ$ period and where different orbital parameters from current values were explored, two separate models designed to spin up the GCM\cite{Toigo2015} were used directly: an energy balance model to determine surface temperature and pressure as well as a 1D atmospheric model. These models agree near the surface with the GCM whenever the full GCM is used. As such, the data shown in Figure 3 are derived from these models whose output was also adjusted in temperature and pressure in the same way as the GCM. 

In addition to the validation of the model against New Horizons data, the values from the GCM give a useful exploration of the diurnal range. The penitente dispersion relation was calculated not only for diurnal average temperatures, but also for the GCM minimum and maximum temperatures, which exhibited a diurnal range of between 4 K and 6 K over the L$_s$ = 0$^\circ$-65$^\circ$ period. 

\subsubsection*{Growth and Age of Plutonian Penitentes}

We use three methods to estimate the age of the penitentes. The lower limit method of applying escape rates\cite{Gladstone2016} and estimates from cratering of nearby terrains\cite{Moore2016} is detailed in the final paragraph of the main text. Therefore, we describe here the range of possible ages based upon the penitente growth rates derived. The penitente amplitude is given by\cite{Claudin2015}:

\begin{equation}
h = h_1\exp(\sigma t)
\end{equation}

\noindent where $h$ is the current height of the features, arising in time $t$ from an initial topographic variation, $h_1$. Such exponential growth rates have been seen both in theoretical models\cite{Claudin2015} and in simulations that directly grow topography through sublimation\cite{Moore2017}. If we are considering only small changes to existing topography, we may derive $\Delta h$, the change over a small interval $\Delta t$ as:

\begin{equation}
\Delta h = h_1 [ \exp(\sigma \Delta t ) - 1]
\end{equation}

\noindent For convenience, we have chosen to provide the value of $t$ in Equations 12 and 13 in terms of orbital cycles of Pluto (Plutonian Years). The value of $\sigma$ as a function of L$_s$, in orbits$^{-1}$, comes from the dispersion relationship (Eq. 1) using the latent heat and density of methane ice of $5.5\times10^5$ J kg$^{-1}$ and 420 kg m$^{-3}$. The deposited flux is also expressed in terms of J m$^{-3}$ orbit$^{-1}$, calculated by summing received energy on a flat surface at 17.5$^\circ$N over a single diurnal cycle and multiplying this value by the number of diurnal cycles per year (\cite{Young1997} 179). Note that this is an instantaneous value and not the total over an orbit.

For convenience, Figure 3 displays a normalized incisement per orbital cycle, $G$, which can be related to the change over a particular time period of:

\begin{equation}
\Delta h=\int G \frac{dt}{\tau}
\end{equation}

\noindent where $\tau$ is the length of one orbital cycle, some 247.94 Earth Years. We select this manner of describing the growth rate to simplify the presentation and accessibility of the figure panel. To give an example of how to interpret these values: a constant value of {$G = 2\,$cm orbit$^{-1}$} over all L$_s$ would integrate to $\Delta h = 2$ cm orbit$^{-1}$.

Equation 14 is used to determine the value of $\Delta h$ per orbit, which can then be used to determine an effective (constant) value for $\sigma$ over all L$_s$ using Equation 13. For 1 cm orbit$^{-1}$, this implies $\sigma = 2\times10^{-5}$ orbit$^{-1}$. Finally, this value is used in Equation 12 to determine the total number of orbits required to produce the observed topography based on assumptions about the scale of the initial topography. For For initial topography 10 m in amplitude, the formation timescale is 50 million years.

\subsubsection*{Code Availability}

The code used to generate the plots shown in this paper and in the extended data set will be available\footnote{www.yorku.ca/jmoores/PlutoPenitentes.tar.gz} upon publication.

\subsubsection*{Data Availability Statement}

The images of Tartarus Dorsa that support the findings of this study can be obtained from the NASA planetary photojournal\footnote{http://photojournal.jpl.nasa.gov} using the PIA identifiers noted in the Figure 1 caption. A map-projected view of the region can be found in reference 5, Figure 3B, and the orientations of the ridges derived from this map-projected view produced by JEM that were used to create the rose diagram of Figure 1a of this study will be available at www.yorku.ca/jmoores/PlutoPenitentes.tar.gz upon publication. All other data used in the production of this paper are listed in the text and the figures were generated from this data using the code described under ‘Code Availability.’

\bibliography{pluto_refs}

\end{document}